\documentstyle[11pt,paspconf,epsf]{article}

\def\lsim{\, \lower2truept\hbox{${< \atop\hbox{\raise4truept\hbox{$\sim$}}}$}\,}
\def\gsim{\, \lower2truept\hbox{${> \atop\hbox{\raise4truept\hbox{$\sim$}}}$}\,}
\begin{document}

\title{Extragalactic Radio Sources and CMB Anisotropies}

\author{L. Toffolatti\altaffilmark{1}, G. De Zotti}
\affil{Osservatorio Astronomico di Padova, vicolo dell'Osservatorio 5,
    35122 Padova, Italy}

\author{F. Arg\"ueso}
\affil{Dpto. de Matem\'aticas, Universidad de Oviedo, c. Calvo Sotelo s/n,
33007 Oviedo, Spain}

\author{C. Burigana}
\affil{Istituto Te.S.R.E., Consiglio Nazionale delle Ricerche, via Gobetti 101,
40129, Bologna, Italy}
% Notice that some of these authors have alternate affiliations, which
% are identified by the \altaffilmark after each name.  The actual alternate
% affiliation information is typeset in footnotes at the bottom of the
% first page, and the text itself is specified in \altaffiltext commands.
% There is a separate \altaffiltext for each alternate affiliation
% indicated above.

\altaffiltext{1}{Dpto. de F\'\i{sica}, Universidad de Oviedo, c. Calvo Sotelo 
s/n,
33007 Oviedo, Spain}

% The abstract is entered in a LaTeX "environment", designated with paired
% \begin{abstract} -- \end{abstract} commands.  Other environments are
% identified by the name in the curly braces.

% Poster authors ONLY may omit the abstract in order to gain a little
% more page space for the text of the poster.

\begin{abstract}
Confusion noise due to extragalactic sources is a
fundamental astrophysical limitation for experiments aimed at accurately
determining the power spectrum of the Cosmic Microwave Background (CMB)
down to arcmin angular scales and with a sensitivity 
$\Delta T/T \simeq 10^{-6}$. At frequencies
$\lsim 200-300$ GHz, the most relevant extragalactic foreground 
hampering the detection of intrinsic CMB anisotropies is constituted by 
radio loud Active Galactic Nuclei (AGN), including
``flat--spectrum'' radiogalaxies, quasars, BL-LACs and blazars.
We review our present understanding of 
astrophysical properties, spectra, and number counts
of the above classes of sources. We also study
the angular power spectrum of fluctuations due both to Poisson distributed
and clustered radio sources and give preliminary predictions
on the power spectrum of their polarized components.
Furthermore, we discuss the capabilities
of future space missions (NASA's MAP, Bennett et al. 1995;
ESA's Planck Surveyor, Bersanelli et al. 1996)
in studying bright radio sources over an almost unexplored frequency
interval where spectral signatures, essential for the understanding of the 
physical processes, show up. 
%All sky surveys spanning the frequency range around the CMB intensity peak
%will be unique in providing complete samples comprising several hundreds
%of radio loud AGN and, possibly, tens of ``inverted''spectrum
%radio sources.

\end{abstract}

% Keywords should be included, but they are not printed in the hardcopy.


% That's it for the front matter.  On to the main body of the paper.
% We'll only put in tutorial remarks at the beginning of each section
% so you can see entire sections together.

\section{Introduction}

In the last fifteen years, deep VLA surveys have allowed to extend
direct determinations of radio
source counts down to $\mu$Jy levels at 1.41, 4.86 and 8.44 GHz. At these 
frequencies, counts now cover about 7 orders of magnitude in flux and 
reach areal densities of several sources arcmin$^{-2}$.

At bright fluxes, the radio source population is dominated by 
classical, stron\-gly evolving, powerful radio galaxies
(Fanaroff-Riley classes I and II) and quasars, whose
counts begin to converge below $\sim 100\,$mJy. The VLA surveys, 
however, have revealed a flattening in differential source counts
(normalized to Euclidean ones) below a few mJy at 1.41 GHz
(Condon \& Mitchell 1984), at 4.86 GHz (Donnelly
et al. 1987; Fomalont et al. 1991), and, most recently, also 
at 8.44 GHz (Windhorst et al. 1993, 1995; Partridge et al. 1997; 
Kellermann et al. 1999; Richards et al. 1998).

Several scenarios have been developed to interpret this ``excess'' 
in the number counts of faint radio sources: 
a non-evolving population of local ($z<0.1$) low-luminosity galaxies
(Wall et al. 1986); strongly evolving normal spirals (Condon 1984, 1989);
 and actively star-forming galaxies (Windhorst et al. 1985, 1987;
Danese et al. 1987; Rowan--Robinson et al. 1993).

Thus, the currently available deep source
counts are more than sensitive  enough to include any radio 
source of the familiar steep and ``flat''-spectrum classes contributing 
to fluctuations detectable by any of the forthcoming space borne 
CMB anisotropy experiments (see Toffolatti et al., 1998;
De Zotti \& Toffolatti, 1998). Extrapolations in flux density are not required:
the real issue is the {\it spectral behaviour} of sources, since existing
surveys extend only up to 8.4 GHz and hence a substantial extrapolation 
in frequency is necessary to reach the frequency bands of the MAP and Planck 
Surveyor missions. The point has to be carefully discussed,
since important spectral features,
carrying information on physical conditions of sources,  
are expected at cm to mm wavelengths. These include the transition from 
optically thick to thin synchrotron emission for ``flat''-spectrum 
sources, the steepening of the synchrotron spectrum due to   
radiation energy losses by the relativistic electrons, and the
mm-wave excesses due to cold dust emission.

On the other hand, future space missions will also provide complete samples 
of the extremely interesting classes of extragalactic radio sources 
characterized by inverted spectra (i.e. flux density increasing with 
frequency), which are very difficult to detect in radio frequency surveys. 
%and are not included in current models (Toffolatti et al., 1995; 1998).
Strongly inverted spectra up to tens of GHz can be produced
in very compact, high electron density regions, by  
synchrotron or free-free absorption.
This is the case for GHz peaked spectrum radio sources 
(GPS), which are currently receiving an increasing amount of interest.
Also of great interest are advection dominated sources (ADS), which 
turn out to have a particularly hard radio emission spectrum.

In \S$\,2$ we briefly discuss the spectral properties, at mm and sub-mm 
wavelengths,   
of the different classes of sources mentioned above.
In \S$\,3$ we deal with number counts while,  
in \S$\,4$, we present estimates of the angular power spectrum of 
intensity and polarization fluctuations 
due to discrete extragalactic sources and discuss the effect of clustering. 
In \S$\,5$ we summarize our main conclusions. 

\section{Radio sources at mm and sub-mm wavelengths}

\subsection{Compact, ``flat''-spectrum radio sources}

\begin{figure}
%\centering
% ****.ps e'il nome del file .ps, width ed height specificano
% le dimensioni. Sono opzionali. Per spostare la figura usare
% \vspace* e \hspace*
%\epsfig{file=FigSED.ps,width=16truecm,height=19truecm}
\plotone{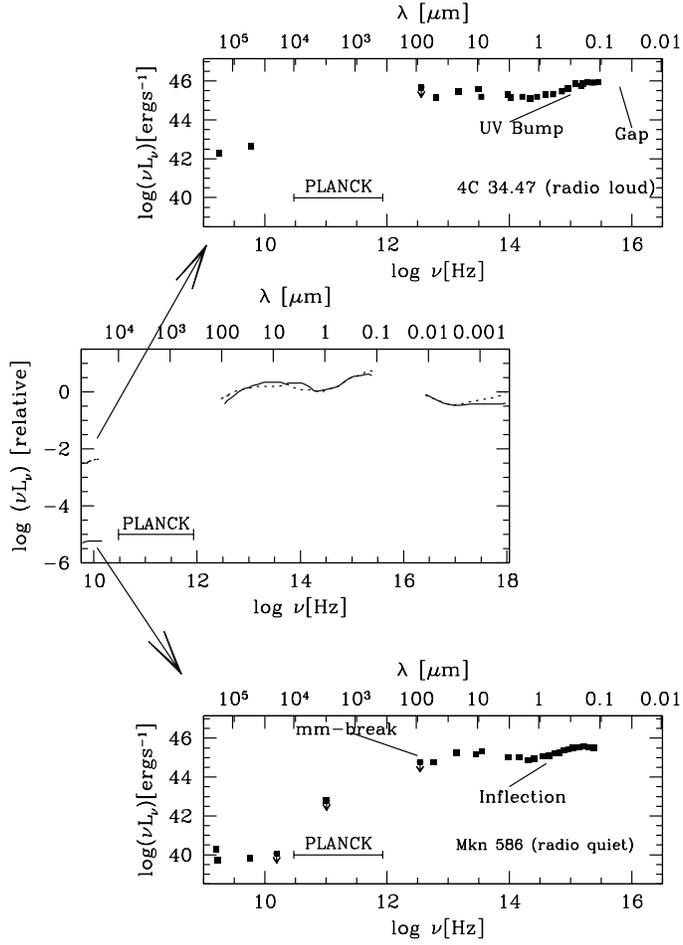}
\vspace{1.truecm}
\caption{Continuum energy distributions for radio-loud and
radio-quiet quasars, adapted from Elvis et al. (1994). The central panel 
shows the mean spectral energy distribution for radio loud ({\it dashed 
line}) and radio quiet ({\it solid line}) quasars, normalized at $1.25\,\mu$m. 
Also shown are the energy distributions of 4C$\,$34.47 ({\it upper panel}) 
and Mkn$\,$586 ({\it lower panel}), representative of the radio-loud 
and radio-quiet class, respectively.}

\label{FigSED}
\end{figure}

The observed spectral energy distributions (SEDs) of   
``flat-''spectrum radio sources (compact radio galaxies, radio loud QSOs,
BL Lacs) generally have a gap at mm/sub-mm wavelengths
(see Figure~\ref{FigSED}). Those sources which have data in this interval  
frequently show a dip in the mm region, indicative of a cross-over of 
two components. 

The spectral shape carries a good deal of extremely interesting
information on the physical properties of sources. 
For example, in flow models of compact radio sources the spectrum 
steepens at the frequency at which the radiative cooling time 
equals the outflow time (cf. Begelman et al. 1984); for ``hot spots'', 
this typically lies in the millimeter or far-IR part of the 
spectrum, while, in cocoons or extended regions of lower surface brightness, 
the break moves down to lower frequencies.   

According to the basic model of Blandford \& Rees (1974) and Scheuer (1974), 
which is supported by a large body of observational evidence, 
the spectral break frequency, $\nu_b$, at which the synchrotron 
spectrum steepens, is related to the magnetic field $B$ 
and to the ``synchrotron age'' $t_s$ (in Myr) by $\nu_b \simeq 96 
(30\,\mu\hbox{G}/B)^{3}t_s^{-2}\,$GHz. Thus, the systematic multifrequency 
study at the Planck and MAP frequencies will provide a 
statistical estimate of the radio source ages and of the evolution of the 
spectrum with cosmic time: both are pieces of information of 
great physical importance. Various evolutionary models of 
the radio emission spectrum have been proposed based on different 
assumptions (``one-shot'' or continuous injection of relativistic electrons,  
complete or no isotropization of the pitch-angle distribution; 
see Myers \& Spangler 1985 for a summary). These models 
strongly differ in the form of the falloff above $\nu_b$; hence 
measurements at mm and sub-mm wavelengths 
will provide crucial information on the physical effects 
operating in radio sources.

Also, many compact ``flat''-spectrum 
sources are observed to become optically thin 
at high radio frequencies. Correspondingly, 
their spectral index steepens to values  
($\alpha \simeq 0.75$) typical of extended, optically thin sources. 

In the case of blazars (Brown et al. 1989) the component dominating at cm
wavelengths is rather ``quiescent'' (variations
normally occur on timescales of years) and has a spectral turnover 
at $\sim 2$--5 cm, where the transition between optically thick and 
optically thin synchrotron emission occurs. At higher frequencies 
the emission is dominated by a violently variable ``flaring'' component, 
which rises and decays on timescales of days to weeks, and has a spectral 
turnover at mm/sub-mm wavelengths. The ``quiescent'' emission may be 
identified with emission from the bulk of a relativistic jet aligned 
close to the line of sight, while the flaring component may arise in 
smaller regions of enhanced emission within the jet, where emitting 
electrons are injected or reaccelerated. The study of the flaring 
component is clearly central to understanding the mechanisms 
reponsible for variability in radio loud active nuclei; the mm/sub-mm 
region is crucial in this respect, since it is in this region that 
the flare spectra become self absorbed.

It is known from VLBI studies that the apparently smooth ``flat'' spectra 
of compact radio sources are in fact the combination of emissions 
from a number of components with varying synchrotron self absorption 
turnover frequencies which are higher for the denser knots. It may be 
argued (Lawrence 1997) that the mm/sub-mm region measures sub-parsec 
scale, unresolved regions, including the radio core.
%; observations in 
%this region are thus particularly important to elucidate the 
%mechanisms operating in these very compact regions. 

Excess far-IR/sub-mm emission, possibly due to dust, is often observed from 
radio galaxies (Knapp \& Patten 1991). Planck data will allow to assess 
whether this is a general property of these sources; 
this would have interesting implications 
for the presence of interstellar matter in the host galaxies, 
generally identified with giant ellipticals, which are 
usually thought to be devoid of interstellar matter.
Observations of large mm fluxes attributed to dust emissions have been 
reported for several distant radio galaxies (see Mazzei \& De Zotti, 1996 
and references therein). The inferred dust masses are 1--2 orders of magnitude 
higher than found for nearby radio galaxies. The two components (synchrotron 
and dust emission) may well have different evolution properties.

\subsection{GHz Peaked Spectrum radio sources}

Current predictions on number counts do not explicitly include 
sources with strongly inverted spectra, peaking at mm wavelengths, 
that would be either missing from, or strongly 
underepresented in low 
frequency surveys and may be difficult to distinguish 
spectrally from fluctuations in the CMB (Crawford et al. 1996).  

The recent study of GHz Peaked Spectrum radio sources (GPS) by O'Dea and Baum 
(1997) has revealed a fairly flat distribution of peak frequencies 
extending out to 15 GHz in the rest frame, suggesting the existence of 
an hitherto unknown population of sources peaking at high frequency 
(see also Lasenby 1996). The host galaxies appear to be a homogeneus
class of giant ellipticals with old stellar populations whereas
GPS quasars present a different redshift distribution and have radio
morphologies quite unlike those of GPS galaxies (Snellen et
al. 1998a,b,c).

It is very hard to guess how common such sources may be. Snellen (1997) 
exploited the sample of de Vries et al. (1997) to estimate a count 
of $22\pm 10\,\hbox{Jy}^{-3/2}\,\hbox{sr}^{-1}$ for sources having 
peak frequencies between 1 and 8 GHz and peak flux densities between 
2 and 6 Jy. He also found that counts of GPS sources are only slowly 
decreasing with increasing peak frequency in that range.
If indeed the distribution of peak
frequencies extends up to several tens GHz
keeping relatively flat, it is conceivable that from several tens to
hundreds of GPS sources will be detected by Planck, whereas
MAP could detect only a few of them.

Therefore, although these rare source will not be a threat for studies 
of CMB anisotropies, we may expect that Planck surveys 
will provide crucial information about their 
properties. GPS sources are important 
because they may be the younger stages of radio source evolution 
(Fanti et al. 1995; Readhead et al. 1996) and may thus provide insight 
into the genesis and evolution of radio sources; alternatively, they 
may be sources which are kept very compact by unusual conditions 
(high density and/or turbulence) in the interstellar medium of the 
host galaxy (van Breugel et al. 1984).

\subsection{Advection-dominated sources}

%Planck/LFI may also allow to study  
Another very interesting class of inverted spectrum radio sources, 
is that characterized by advection-dominated emission 
(Narayan \& Yi 1995; Fabian \& Rees 1995; Di Matteo \& Fabian 1997). 
Convincing observational evidence for the presence of super-massive 
black-holes (BH) in many nearby galaxies 
have been accumulating in recent years 
(Kormendy \& Richstone 1995; Magorrian et al. 1998; Ho 1999). These
data seem to imply that an important, possibly dominant,
fraction of all massive galaxies
with appreciable spheroidal component (hence the E/S0 galaxies in 
particular, but also early-type spirals) host a black-hole with a mass 
roughly proportional to that of the hosting spheroid. Franceschini et 
al. (1998) have found a tight relationship between the BH mass and
both the nuclear and total radio flux at centimetric wavelengths. 
The radio flux turns out to be proportional to $M_{\rm BH}^{2.5-3}$. 
This is
consistent with the radio centimetric flux being contributed by
cyclo-synchrotron emission in an advection-dominated accretion flow.
The latter should correspond to a situation in which the accretion rate 
is low, as typically expected for the low-density ISM in local early
type galaxies.

A property that distinguishes this emission is an inverted spectrum
with spectral index $\alpha = 0.4$ up to a frequency of 100--200 GHz,
followed by fast convergence (far-IR and optical emission is 
expected to be very weak). Based on the analysis by Franceschini 
et al. (1998), we would expect that some  
sources of this kind may detected by Planck 
at 70 and 100 GHz. This assumes that the advection flows evolves
in redshift $\propto (1+z)^3$, as suggested by analyses of
faint radio-optical samples of E/S0's. In spite of the limited 
statistics, this would be a way to test for the presence of 
massive BH's and of the evolution of the ISM in galaxies up to
moderate redshifts.

\subsection{Free-free self absorption cutoffs in AGN}

High frequency free-free  cutoffs may be present in AGN
spectra. Ionized gas in the nuclear region absorbs radio photons 
up to a frequency:

\begin{equation}
\nu_{\rm ff} \simeq 50 {g \over 5} {n_e \over 10^5\,{\rm cm}^{-3}}
\left({{\rm T}\over 10^4\,{\rm K}}\right)^{-3/4} l_{\rm pc}^{1/2}\ {\rm
GHz}\ .
\end{equation}
Free-free absorption cutoffs at frequencies $> 10\,$GHz may indeed be
expected, in the framework of the standard torus scenario for
type 1 and type 2 AGN, for radio cores seen edge on, and may have been
observed in some cases (Barvainis \& Lonsdale, 1998). They
provide constraints on physical conditions in the parsec scale accretion
disk or infall region for the nearest sources of this kind.

\section{Counts of radio sources at cm to mm wavelenghts}

Source counts are presently available only at cm wavelengths. 
In carrying out extrapolations to mm wavelengths, several effects need to be 
taken into account. On one side, 
the majority of sources with flat or inverted spectra at 5 GHz 
have spectral turnovers below 90 GHz (Kellermann \& Pauliny-Toth 1971; 
Owen \& Mufson 1977). This is not surprising since, as noted above, 
astrophysical 
processes work to steepen the high frequency source spectra. 

On the other side, 
high frequency surveys preferentially select sources with 
harder spectra. For power law differential source counts, $n(S,\nu_0) = 
k_0\,S^{-\gamma}$, and a Gaussian spectral index distribution with 
mean $<\alpha>_0$ and dispersion $\sigma$, the counts at a frequency $\nu$ 
are given by $n(S,\nu) = n(S,\nu_0)(\nu/\nu_0)^{\alpha_{\rm eff}}$ 
with (Kellermann 1964; Condon 1984):
$\alpha_{\rm eff}= <\alpha>_0 + \ln(\nu/\nu_0)\sigma^2(1-\gamma)^2/2$. 
Estimates neglecting the dispersion of spectral indices underestimate 
the counts by a factor $\exp[\ln^2(\nu/\nu_0)\sigma^2(1-\gamma)^2/2]$.
The spectral index distribution between 5 and 90 GHz determined by Holdaway et 
al. (1994) has $\sigma = 0.34$; for Euclidean counts, $\gamma =2.5$, 
the correction then amounts to about a factor of 3.

A good fraction of the observed spread of spectral indices is due to 
variability whose  rms amplitude, in the case of blazars,  
increases with frequency, reaching a factor of about 1.5 at a few hundred GHz
(Impey \& Neugebauer 1988). 
In some cases, variations by a factor of 2 to 3, or more, have been observed  
(e.g. 3C345: Stevens et al. 1996; PKS$\,0528+134$: Zhang et al. 
1994; $0738+545$ and 3C279: Reich et al. 1998). 
The highest frequency outbursts are expected to be associated to the 
earliest phases of the flare evolution. Since the rise of the flare is often 
rather abrupt (timescale of weeks), they were probably frequently missed.  

In view of the uncertainties on the spectra of radio selected AGNs,
and the poor knowledge on number counts and spectral properties of
inverted spectrum sources discussed in \S{2},
an accurate modelling of radio source counts at $\nu\sim 100$ GHz
is currently impossible. However, the simple model adopted by 
Toffolatti et al. (1998) appears to be remarkably successful.

\begin{figure}
\plotone{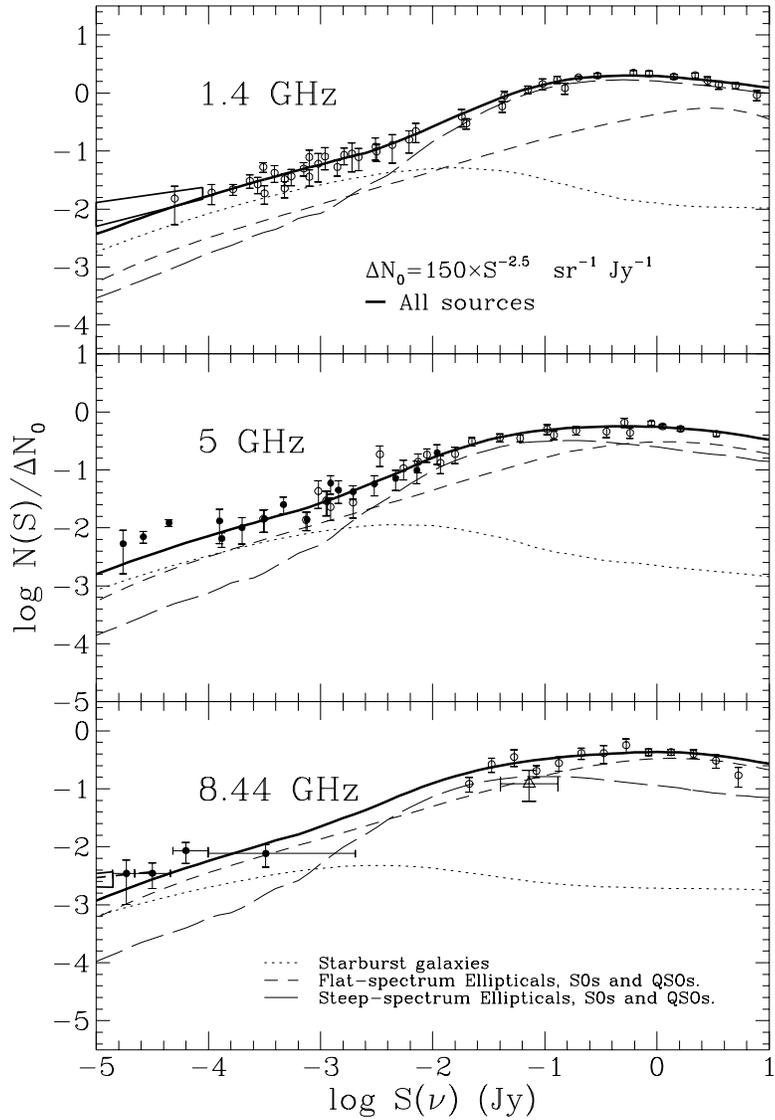}
\vspace{1.truecm}
\caption{ 
Comparison between predicted and observed differential source 
counts at 1.4, 5 and 8.44 GHz normalized to
$150 S^{-2.5}\,\hbox{sr}^{-1}\,\hbox{Jy}^{-1}$ (see top panel),
adapted from Toffolatti et al. (1998).
The contributions of the most relevant classes
of radio sources according to the model of Danese et al. (1987) are shown.
For references on the data points see Danese et al. (1987);
additional data are from Donnelly et al. 
(1987), Fomalont et al. (1991), Windhorst et al. (1993), Partridge et al.
(1997).}
\end{figure}

These authors adopted the luminosity evolution
scheme of Danese et al. (1987), who considered three classes of sources: 
powerful radio galaxies and radio loud quasars (distinguishing between 
extended, steep spectrum, and compact, ``flat"" spectrum sources), and 
evolving starburst/interacting galaxies.  
The average spectral index, $\alpha$
($S\propto\nu^{-\alpha}$), of ``flat--spectrum'' sources was taken to be 
$\alpha=0.0$ for $\nu \leq 200$ GHz,
with a steepening to $\alpha =0.75$ at higher frequencies.
As for ``steep''--spectrum sources (elliptical, S0 and starburst
galaxies), whose contribution to source counts is actually minor in the
whole frequency range of interest in connection with the MAP and Planck 
missions, the radio power--spectral index
relation determined by Peacock and Gull (1981) was adopted.
This simple recipe has allowed to reproduce,
without any adjustment of the parameters,
the deep counts at 8.44 GHz (Windhorst et al. 1993; Partridge et al. 1997), 
which were produced several years after the model (see Figure 2). 

This scenario, implying a substantial contribution of active star-form\-ing  
galaxies to sub-mJy counts at cm wavelengths, 
is consistent with the results  
by Kron et al. (1985) and Thuan \& Condon (1987) indicating that 
the optical counterparts of sub-mJy sources are mainly 
faint blue galaxies, often showing
peculiar optical morphologies indicative of high star formation activity
and of interaction and merging phenomena. Moreover, the spectra are
similar to those of the star--forming galaxies detected by IRAS
(Franceschini et al. 1988; Benn et al. 1993).
On the other hand, a recent work by Gruppioni, Mignoli \& Zamorani
(1999), reaching a fainter magnitude limit in the spectroscopic
identifications of sources selected at 1.4 GHz with a flux limit 
of 0.2 mJy, finds that the majority of their radio sources are 
likely to be early type galaxies.

Optical identifications of sources at $\mu$Jy flux levels (Hammer et al. 1995;
Windhorst et al., 1995; Richards et al., 1998) show that they 
are made mainly of star-forming galaxies, with 
a smaller fraction of ellipticals, late-type galaxies with nuclear 
activity and local spiral galaxies. 

It may be noted that the counts predicted by Toffolatti et al. (1998) at
frequencies above 100 GHz may be somewhat depressed by the assumption 
of a spectral break at 200 GHz for all ``flat''-spectrum sources 
while examples are known of sources with a flat or 
inverted spectrum up to 1000 GHz. In view of this, we updated the
the baseline model of Toffolatti et al. by assuming that the
steepening of the spectral indices to $\alpha =0.75$ occurs at
$\nu =1000$ GHz.

\subsection{Comparison with other estimates of source counts}

Sokasian et al. (1999) have produced skymaps of bright radio sources at
frequencies up to 300 GHz by means of detailed individual fits of the
spectra
of a large number of sources compiled from several catalogs, including
the all sky sample of sources with $S_{5\,{\rm GHz}} \geq 1\,$Jy
(K\"uhr et al. 1981). Their estimated number of sources
brighter than $S_{90\,{\rm GHz}} = 0.4\,$Jy is about a factor of two
below that predicted by Toffolatti et al. (1998). A very similar result
was obtained by Holdaway et al. (1994) based on the observed distribution
of 8.4--$90\,$GHz spectral indices. On the other hand, these empirical
estimates yield, strictly
speaking, lower limits, since sources with inverted spectra may
be under-represented in the primary sample. Furthermore, in the presence of
substantial variability, estimates using mean fluxes underpredict 
actual counts of bright sources.

A comparison of the $\log N$--$\log S$ of radio sources
with the predicted ones for dusty galaxies (even allowing for very different
evolutionary scenarios: De Zotti et al. 1999, their Figure 1)
shows an abrupt change in the 
populations of bright sources observed in channels above and below
$\sim 1\,$mm: radio sources dominate at the longer wavelengths and
dusty galaxies at the shorter ones. This is due to the steep increase
with frequency of the dust emission spectrum
in the mm/sub-mm region (typically $S_\nu \propto \nu^{3.5}$), which makes
the crossover between radio and dust emission components
only weakly dependent on their relative intensities; moreover,
 dust temperatures tend to be higher for distant high luminosity
sources, partially compensating for the effect of redshift. 

At, say, $\nu\geq 200$--300 GHz dusty galaxies dominate
the number counts of extragalactic sources. 
We only briefly remind the point in the next
subsection and defer to the comprehensive reviews of Guiderdoni et al. (1999) 
and of Mann et al. (1999) for a thorough discussion.

\subsection{Counts of dusty galaxies}

Although the situation is rapidly improving, thanks to the
deep ISO counts at $175\,\mu$m (Kawara et al. 1997; Puget et al. 1999), to  
the preliminary counts at $850\,\mu$m with SCUBA on JCMT 
%Smail et al. 1997; Hughes et al. 1998; Barger et al. 1998; Eales et al. 1998;
%Blain, et al., 1998;
(see Mann et al. 1999, for a review on the subject)
and to the important constraints from measurements of the far-IR to 
mm extragalactic background (Schlegel et al. 1998; Hauser et al. 1998; 
Fixsen et al. 1998), 
first detected by Puget et al. (1996), current estimates are affected by 
bigger uncertainties than in the case of radio sources.

In fact, predicted counts have a higher responsiveness to the poorly known 
evolutionary properties, because of the 
boosting effect of the strongly negative K corrections.  The most extensive 
surveys, carried out by IRAS at $60\,\mu$m, cover a limited range in flux 
and are rather uncertain at the faint end (Hacking \& Houck 1987; 
Gregorich et al. 1995; Bertin et al. 1997). It is then not surprising that 
predictions of recent models differ by substantial factors.

Again, substantial extrapolations in frequency are required, and have 
to deal with the poor knowledge of the spectrum of galaxies in the mm/sub-mm 
region; the $1.3\,$mm/$60\,\mu$m flux ratios of galaxies are observed 
to span about a factor of 10 (Chini et al. 1995; Franceschini \& 
Andreani 1995).

\section{Angular power spectrum of source fluctuations}

%The angular power spectrum of Galactic emission, however, falls rapidly at 
%small angular scales ($C_\ell \propto \ell^{-3}$: Gautier et al. 1992, Kogut 
%1996, Wright 1998; somewhat flatter slopes have been 
%reported for some regions of the sky, cf. Lasenby 1996, Schlegel et al. 1998), 
%while a Poisson distribution of extragalactic point sources produces a 
%white-noise power spectrum, with the same power on all multipoles (Tegmark \& 
%Efstathiou 1996). The estimates by Toffolatti et al. (1998) indicate that, at 
%high galactic latitude, foreground fluctuations are dominated by extragalactic 
%sources for scales $\lsim 30'$. The minimum moves to somewhat higher 
%frequencies. Our calculations indicate that, for fluctuations in a beam of 
%width $\theta \lsim 30'$ the minimum is close to $100\,$GHz 
%while the minimum in the temperature power spectrum for the corresponding 
%multipoles ($\ell > 300$) is close to 
%$150\,$GHz.

The effect of radio sources as a limiting factor
for the detection of primordial CMB anisotropies has been
extensively analyzed by many authors (see, e.g., Franceschini et al. 1989;
Blain \& Longair 1993; Danese et al. 1996; Gawiser \& Smoot 1997).
More recently, detailed analyses of the problem have been worked out 
by Toffolatti et al. (1998) for the Planck Surveyor mission, and by
Refregier, Spergel \& Herbig (1999) for the MAP mission.

The relevant formalism is readily available in the literature
(see, e.g., De Zotti et al. 1996; Tegmark \& Efstathiou 1996; Toffolatti
et al. 1998; Scott \& White 1999) and we skip it here.
A Poisson distribution of extragalactic point sources produces a simple
white-noise power spectrum, with the same power in all multipoles, 
so that their contribution to fluctuations in a unit logarithmic 
multipole interval increases with $\ell$ as $\ell(\ell +1)C_\ell \propto 
\ell^2$ (for large values of $\ell$),  
while, at least for the standard inflationary models, 
which are consistent with the available anisotropy detections,  
the function $\ell(\ell +1)C_\ell$ yielded by primordial CMB fluctuations 
is approximately constant for $\ell \lsim 100$, then oscillates and finally 
decreases quasi exponentially for $\ell \gsim 1000$  
($\theta \lsim 10'$).
Hence confusion noise due to discrete sources will 
dominate at small enough angular scales.

Figure 3 shows the expected angular power spectra of the extragalactic
foregrounds components at 30, 100 and 217 GHz, corresponding to the 
central frequencies of three Planck channels; also, MAP has 
channels centered at 30 and 94 GHz.
%Moreover, at $\nu\geq 200-300$ GHz dusty
%galaxies dominates the number counts of extragalactic source
%and thus radio sources are not the dominant contributors to sky fluctuations.
%Moreover, these three frequencies also correspond to three Planck channels
%at which CMB polarization fluctuations will be measured (see \S{5.2}).
The two lines for each population (radio sources and far-IR sources) 
are obtained assuming that  sources can be identified and removed 
down to fluxes of 1 or 0.1 Jy.

Radio sources give the most relevant
contribution to CMB fluctuations up to, say, $\sim 200$ GHz,
whereas dusty galaxies dominate at higher frequencies.
At 217 GHz, the Poisson fluctuation level due to far-IR
selected sources predicted by the model of Guiderdoni et al. (1998)
is higher than that of radio selected sources, whereas it falls below
the radio component if we adopt the Toffolatti et al. model. As 
discussed by De Zotti et al. (1999), the latter model falls somewhat 
short of the best current estimate of the far-IR to mm extragalactic 
background (Fixsen et al. 1998), while model E by Guiderdoni et al. tends to 
exceed it. 

Figure 3 also indicates that source removal is much more
effective in reducing the Poisson fluctuation level at $\lambda \gsim 1\,$mm.
In fact, in this wavelength range fluctuations are dominated
by the brightest sources below the detection
limit, while at shorter wavelengths the dominant population are evolving
dusty galaxies whose counts are so steep that a major contribution
to fluctuations comes from fainter fluxes.

\begin{figure}
\vskip-10pt
\plotone{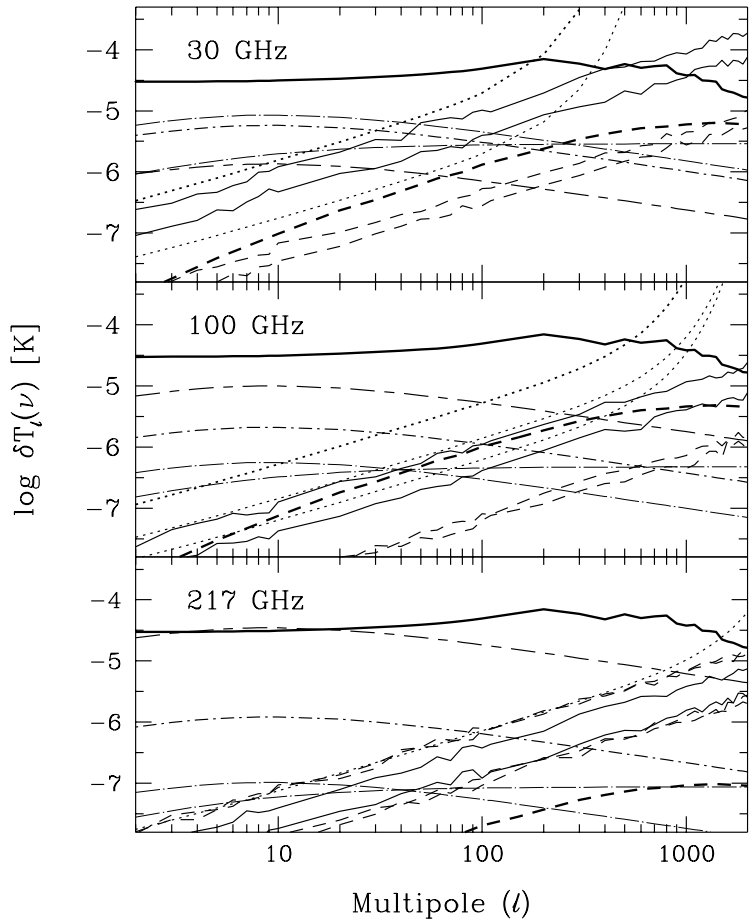}
\vskip-20pt
\caption{Angular power spectra of the foreground components contributing to
fluctuations at 30, 100 and 217 GHz. Following Tegmargk \& Efstathiou
(1996), we have
plotted, for each component, the quantity $\delta T_\ell(\nu) =
[\ell(2\ell+1)C_\ell(\nu)/4\pi]^{1/2}$.
The upper heavy solid curve shows the power spectrum of CMB
fluctuations predicted by the standard CDM model ($\Omega=1$,
$H_0= 50\,{\rm km}\,{\rm s}^{-1}\,{\rm Mpc}^{-1}$, $\Omega_b  =
0.05$). The dotted lines show the unsmoothed noise contributions
of the MAP (upper line) and Planck instruments; at 100 GHz both the HFI 
and LFI (lowest line) noise curves are plotted. The roughly diagonal
thin lines show the contributions of extragalactic sources
(radio sources: solid lines; far-IR sources: dashed lines). At 217 GHz
the prediction of Guiderdoni et al. 1998 -- their model E -- is also
plotted. We assume that sources brighter than 1 and 0.1 Jy can be
identified and removed and we neglect the effect of clustering (see 
\S{4.1}). The heavy dashed lines, flattening at large $\ell$, refer to 
fluctuations due to the Sunyaev-Zeldovich effect. 
The angular power spectra of the
galactic synchrotron, free-free and dust emission are also plotted.
See text for more details.}
\end{figure}

Fluctuations in the Galactic emissions at $|b|>30^\circ$
are also plotted in Figure 3, for comparison.
They are represented
by long$+$short dashes (dust), dots$+$short dashes  (free-free), and
dots$+$long dashes [synchrotron: we have plotted both the power spectrum 
derived by Tegmark \& Efstathiou (1996), $C_{\ell}\propto \ell^{-3}$, 
and that observed in the Tenerife patch, $C_{\ell}\propto \ell^{-2}$  
(Lasenby 1996); 
the latter is significantly lower than the former in the range of scales
where it has actually been measured, but has a flatter slope so that it
may become relatively more important on small scales].
The heavy dashed line flattening at large $\ell$ shows the power
spectrum of anisotropies due to the Sunyaev-Zeldovich effect computed
by Atrio-Barandela \& M\"ucket (1998)
adopting a lower limit of $10^{14}\,{\rm M}_\odot$ for
cluster masses, a present ratio $r_{\rm virial}/r_{\rm core}= 10$, and
$\epsilon = 0$.

For $\ell < 300$, corresponding to angular scales $> 30'$
[$\ell \simeq 180^\circ/\theta({\rm deg})$], diffuse
Galactic emissions dominate total foreground fluctuations even at high Galactic
latitudes. These are minimum at $\nu \simeq 70\,$GHz (Kogut 1996).
For larger values of $\ell$ the dominant contribution is from extragalactic
sources; their minimum contribution to the
anisotropy signal occurs around 150-200$\,$GHz.
Therefore, the minimum in the global power spectrum of foreground
fluctuations moves, at high galactic latitudes, from about 70 GHz for
$\ell < 300$, to 150-200$\,$GHz at higher values of $\ell$.
The detailed spectral and evolutionary behaviour of sources determines
the exact value of the frequency of the minimum foreground fluctuations; 
such frequency decreases with decreasing Galactic latitude (De Zotti et al.
1999).

On the other hand, in the frequency range 50--200 GHz, a higher
contamination level due to source confusion could be produced by a
still undetected population of sources whose emission peaks at
$\sim 100$ GHz (Gawiser, Jaffe \& Silk, 1999). Anyway, 
the discussion in \S{1} and \S{2} indicates that {\it unknown} source
populations are unlikely to give a fluctuation level much higher
than the presently estimated one.

\subsection{The effect  of clustering}

Toffolatti et al. (1998)
found that clustering contribution to fluctuations
due to extragalactic sources is generally small in comparison with the
Poisson contribution.
However, the latter, in the case of
radio sources, comes mostly from the brightest sources below the detection
limit, while the clustering term is dominated by fainter sources. Therefore,
an efficient subtraction of radio sources decreases the Poisson term much more
effectively than the clustering term, which therefore becomes relatively
more and more important.

In the case of a power law angular correlation function ($w(\theta) \propto
\theta^{1-\gamma}$), the power spectrum of intensity fluctuations is
$C_\ell \propto \ell^{\gamma -3}\ $ (Peebles 1980; eq. 58.13).
If this behaviour extends to large enough angular scales,
i.e. to small enough values of $\ell$, the clustering 
signal, for a power law index not much steeper than the canonical value 
$\gamma \simeq 1.8$, 
will ultimately become larger than the Poisson anisotropy. On the
other hand, for large values of $\theta$, $w(\theta)$ is expected to drop
below the above power law approximation, and $C_\ell$ will
correspondingly break down.

\begin{figure}[t!] % fig 4
%\centerline{\epsfig{file=got.ps}}
%\vspace*{-30pt}
%\centerline{\epsfig{file=pslficlu.ps,height=9truecm,width=15truecm}}
\plotone{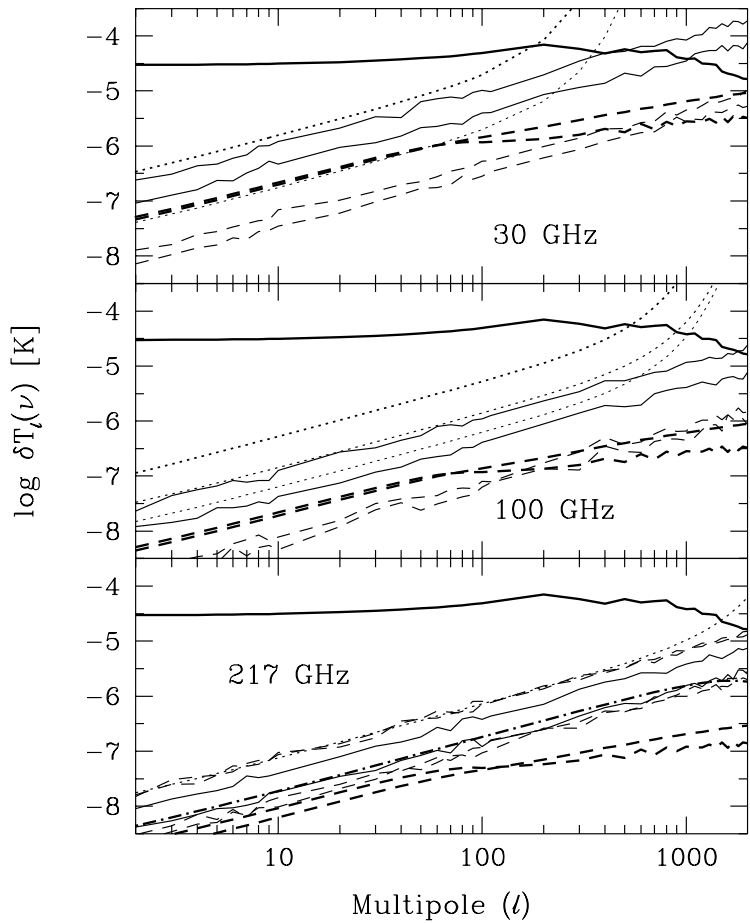}
\vspace{1.truecm}
\caption{Comparison of the Poisson component of the power spectra of
radio and far-IR selected sources with the component due to clustering.
The thick solid and the (thick and thin) dotted lines have the same meaning
as in Figure 3.
The roughly diagonal solid and dashed thin lines represent the Poisson
component due to radio and far-IR selected sources, respectively
(with the same limits for detection and removal in as Figure 3).
The heavy dashed lines show the component due to clustering of radio sources
estimated using
the angular correlation functions derived by Loan et al. (1997; lower line)
and by Magliocchetti et al. (1999). The heavy dot/dashed line
represents the contribution due to clustering of far-IR sources
estimated based on the angular correlation function of Lyman break galaxies.
}
\end{figure}

The preliminary estimates (Toffolatti et al., in
preparation) reported in Figure 4 are based on the
correlation functions of radio sources
derived by Loan et al. (1997) and by Magliocchetti et al. (1999).
The different slopes at large values of $\ell$ reflect
different values of $\gamma$: the upper curve correspond to the larger
value ($\gamma = 2.5$) obtained by Magliocchetti et al. (1999).

Scott \& White (1999) have recently shown that if dusty galaxies
cluster like the $z \sim 3$ Lyman break galaxies (Giavalisco et al. 1998),
 at frequencies $\geq 217\,$GHz
the anisotropies due to clustering may exceed the Poisson ones on all scales
accessible to Planck; in the 353$\,$GHz ($850\,\mu$m) channel the clustering
signal may exceed the primordial CMB anisotropies on scales smaller than
about $30'$. It should be noted, however, that current models (Toffolatti
et al. 1998; Guiderdoni et al. 1998) strongly suggest a broad redshift
distribution of sources contributing to the autocorrelation function of
the intensity fluctuations, implying a strong dilution of the
clustering signal. A further substantial overestimate of the effect of
clustering may follow from the extrapolation to degree scales, with
constant slope, of the angular correlation function determined on
scales of up to a few arcmin. Our preliminary estimate shown in 
Figure 4, has been obtained by assuming that  
the angular correlation function of Lyman break galaxies substantially 
steepens at $\theta\simeq 6'$, consistent with the 
data of Giavalisco et al. (1998).

\subsection{Polarization}

The polarized spectra have not been studied as extensively as the
intensity spectra. For a power law electron energy spectrum $dN/dE=N_0
E^{-p}$, the polarization level, $\Pi$, yielded by a
uniform magnetic field is $\Pi = 3(p +1)/(3p+7)$ (LeRoux 1961).
For a typical high frequency value of $p \sim 3$, $\Pi\sim 75\%$.
Non-uniformities of the magnetic fields and differential
Faraday rotation decrease the
polarization level. However, the Faraday rotation optical depth is
proportional to $\nu^{-2}$ so that Faraday depolarization is negligible
at the high frequencies relevant for Planck and MAP.

%The intrinsic polarization of {\it Galactic dust emission} is estimated to
%be
%$\sim 30\%$ \cite{Hildebrand}. The observed polarization degree
%is smaller by a factor $\Phi =RF\cos^{2}\gamma$ \cite{Lee}, where
%$R$ and $F$ are the reduction factors due to misalignments of grain axes
%with the magnetic field and to the different orientations of polarization
%vectors of different components along any line of sight, while the
%$\cos^2\gamma$ accounts for the projection of the direction of polarization
%on the plane of the sky.

%The level of polarized emission from Galactic dust
%at high Galactic latitudes has been estimated by \cite{Prunet}.
%To ease
%the comparison with the CMB polarization, they considered two linear
%combinations of the Stoke's parameters in Fourier space: the E-mode,
%dominated
%by scalar perturbations, and the B-mode, produced by tensor perturbations.
%For Galactic dust polarized emission, the power spectra of the two
%modes can be approximated by \cite{Prunet}:
%
%\begin{equation}
%C_E(\ell) = 8.9\times 10^{-4} \ell^{-1.3}\ (\mu{\rm K})^2
%\end{equation}
%
%\begin{equation}
%C_B(\ell) = 1.\times 10^{-3} \ell^{-1.4}\ (\mu{\rm K})^2
%\end{equation}
%
%They concluded that, at frequencies around 100 GHz, the power spectrum of
%this polarized component is below that of the scalar induced CMB
%polarization
%at $\ell \gsim 200$, but is above the tensor induced CMB polarization
%expected for a flat, tilted cold dark matter model with spectral index of
%scalar perturbations $n_s=0.9$.

Bouchet et al. (1998) have analized the possibility of extracting the power
spectrum of CMB polarization fluctuations in the presence of polarized
Galactic foregrounds using a multifrequency Wiener filtering of the data.
They concluded that the power spectrum of E-mode polarization of the CMB
can be extracted from Planck data
with fractional errors $\lsim 10$--30\% for
$50 \lsim \ell \lsim 1000$.
The B-mode CMB polarization, whose detection
would unambiguously establish the presence of tensor perturbations
(primordial gravitational waves),
can be detected by Planck with signal-to-noise
$\simeq 2$--4 for $20 \leq \ell \leq 100$ by averaging over a 20\%
logarithmic range in multipoles.

\begin{figure}
\plotone{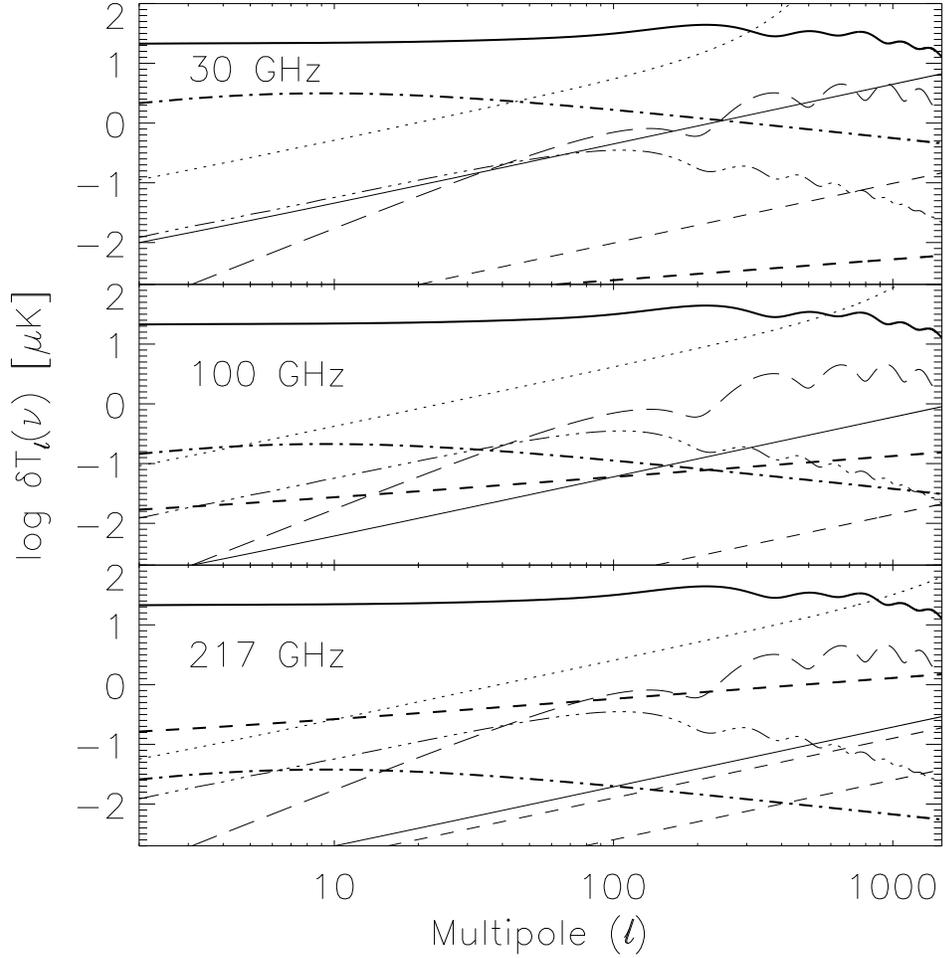}
\vspace{1.truecm}
\caption{Power spectrum of polarization fluctuations at 30, 100 and 217 
GHz. The thick solid lines show the angular power spectrum
of the CMB temperature anisotropies
whereas the long dashes (three dots-dashed) lines show
the power spectrum of polarized scalar (tensor) CMB fluctuations,
respectively. A tilted CDM model with standard parameters and 
scalar to tensor quadrupole ratio $7(1-n_s)$, $n_s=0.9$, $n_t=0.1$, has been 
adopted.  The dotted lines show 
the instrumental noise power spectrum for polarization measurements.
The thin solid (thin dashed)
lines represent estimates of the angular power
spectrum of extragalactic radio (far-IR) source fluctuations assuming that
the emission of radio and far-IR sources is polarized, on average,
at the 5\% and 2\% level, respectively.
Estimates of the angular power spectra of the polarized synchrotron
(thick dot-dashed line) and dust (thick dashes) emission of the Galaxy
are also shown (see text).
}

\end{figure}

Polarization of extragalactic radio sources,
not considered by Bouchet et al. (1998),
is an important issue as well.
Flat spectrum radio sources are typically 4-7\% polarized at cm
and mm wavelengths (Nartallo et al. 1998; Aller et al. 1999).
For random orientations of the magnetic
fields of sources along the field of view, the rms polarization
fluctuations are approximately equal to intensity fluctuations times
the mean polarization degree (De Zotti et al., in preparation).

Our preliminary estimate of Figure 5 assumes a mean polarization of 
radio sources of 5\%, close to the main peak of the percentage 
polarization distribution of the sample studied by Nartallo et al. (1998). 

Measurements of polarized thermal emission from dust 
are only available for interstellar clouds in our own Galaxy. 
The distribution of observed polarization degrees of dense clouds 
at 100$\,\mu$m shows a peak at $\sim 2\%$ (Hildebrand 1996). 
We have adopted this value for mean polarization of dusty galaxies 
at mm/sub-mm wavelengths; this is likely to be an upper limit 
since the global percentage polarization from a galaxy is the average of 
contributions from regions 
with different polarizing efficiencies and different orientations of 
the magnetic field with respect to the plane of the sky; all this works to 
decrease the polarization level in comparison with the mean of individual 
clouds. 

At 217 GHz the power spectrum obtained on the basis of the source counts
of Guiderdoni et al. (model E) is also shown (upper thin dashed line).
Note that, since the emission of radio sources appears to be more 
polarized than that of dusty galaxies, radio sources give the dominant 
contribution to extragalactic polarization fluctuations up to 217 GHz. 
Anyway, Figure 5 shows that the main limitations to CMB polarization 
measurements come from Galactic emissions, not from extragalactic sources.

A brief summary on polarization fluctuations of Galactic emissions  
can be found in De Zotti et al. (1999). 
The power spectrum of the polarized component of 
Galactic synchrotron emission, shown in Fig. 5, 
has been obtained from the corresponding temperature fluctuation
power spectrum (as estimated by Tegmark \& Efstathiou 1996)
assuming the signal to be 40\% polarized. As for dust emission, we 
have taken up the estimates by 
Prunet et al. (1998) for the E-mode; the extrapolations to 
to 100 and 30 GHz are made adopting their  
dust emission spectrum.
Free-free emission is not polarized. However, Thompson scattering by
electrons in the HII regions where it is produced, may polarize it
tangentially to the edges of the electron cloud (Keating et al. 1998).
The polarization level is expected to be small, with an upper limit of
approximately 10\% for an optically thick cloud.

%\section{Observations}

                                         %% beginning of font "text"
\section{Conclusions}

The very deep VLA surveys
have allowed to extend radio source counts down to $\mu$Jy levels at $\lambda
\geq 3$ cm; the bulk of radio sources in the Universe have probably been  
detected (Haarsma \& Partridge 1998). On the other hand, estimates of 
the counts and of fluctuations due to extragalactic sources at mm/sub-mm 
wavelengths are made uncertain by the poor knowledge of their spectral 
properties in this spectral region as well as by the possibility that 
source populations with strongly inverted spectra may show up. 
However, as mentioned in \S{3}, estimates based on very different approaches 
agree to within a factor of 2 up to $\simeq 100\,$GHz.
 
Conservative estimates indicate 
that, in the frequency range 50--200 GHz, extragalactic foreground
fluctuations are well below the expected amplitude of CMB
fluctuations on all angular scales covered by the Planck and MAP missions.

Current data on the angular correlation function of radio sources imply 
that fluctuations due to clustering are generally small in comparison
with Poisson fluctuations; however, the relative importance of the  former 
increases 
if sources are subtracted from the sky maps down to faint flux levels.

Fluctuations due to clustering may be more important 
at frequencies higher than $\nu\sim 200-300$ GHz, if high 
redshift dusty galaxies cluster like the Lyman break galaxies at $z \sim 3$. 

The polarized emission of extragalactic sources will not be
a threat for measurements of the polarized component of primordial
CMB anisotropies; much stronger limitations come from Galactic synchrotron 
and dust emissions.

Moreover, future space mission like Planck (and, to a lesser extent, MAP)
will not only provide all sky maps of the primordial CMB anisotropies
but also unique information on the physics of
compact radio sources and in particular 
on the physical conditions in the most compact components (transition 
from optically thick to optically thin synchrotron emission, ageing of 
relativistic electrons, high frequency flares) and on their relationship 
with emissions at higher energies (SSC versus EC models). They 
will also allow us to study  
the population properties of inverted spectrum sources like 
GPS, sources with high frequency free-free self absorption, ADS, etc..

% In this section, we see the use of the \subsection command to set off
% an independent subsection.  We only have one here; usually there would
% be several.
%
% We show the use of several of the displayed math environments described
% in the User Guide, and you get a healthy dose of mathematical typesetting
% examples.  Also, observe the use of the LaTeX \label command after the
% \subsection to give a symbolic tag to the subsection for cross-referencing
% in a \ref command.  LaTeX automatically numbers the sections, equations,
% tables, etc. as it goes, so in general you don't know what number something
% is going to have.  We'll refer to the "hairymath" section a little later.

\vspace{1.truecm}

\acknowledgments

LT thanks the organizing committee and the Sloan Foundation for their warm 
hospitality. We thank B. Guiderdoni for kindly providing us with the source
counts predicted by his model E. The angular power spectrum of CMB
anisotropies has been calculated by CMBFAST 2.4.1 (Seljak U., \&
Zaldarriaga M., 1996). We gratefully acknowledge the long-standing,
very fruitful collaboration on extragalactic foregrounds with L. Danese,
A. Franceschini and P. Mazzei. We thank N. Mandolesi for useful discussions
on the capabilities of the Planck Surveyor mission and G.L. Granato
for kindly providing us with Figure 1.
LT also thanks F. Bouchet, R.D. Davies, E. Gawiser, E. Guerra, B. Partridge,
A. Refregier and D. Scott for helpful discussions and comments during
his stay in Princeton. This research has been supported in part by
grants from ASI and CNR. LT acknowledges partial financial support from the 
Spanish DGES, project PB95--1132--C02--02, and Spanish CICYT Acci\'on
Especial n. ESP98-1545-E.

% That's the end of the main body of the paper.  Now we will have some
% back matter.

% Now comes the reference list.  Since we typed out the citations ourselves,
% the reference list is enclosed in a "references" environment.  Each
% new reference begins with a \reference command which sets up the proper
% indentation.  Typography that may be required in the reference list by
% the editorial staff must be included by the author.
%
% Observe the "standard" order for bibliographic material: author name(s),
% publication year, journal name, volume, and page number for articles.
% Some journal names are available as macros; see the WGAS markup
% instructions for a listing of which ones have been "macro-ized".
% Note the use of curly braces to delimit the font changes: it is essential
% that this be done to limit the scope of the font declaration.
%
% There is no need to engage in any other typographic manipulation.

\end{document}